\documentclass[prl,amsmath,amssymb,twocolumn,superscriptaddress,showpacs]{revtex4}
\usepackage{amsmath,amssymb}
\usepackage[usenames]{color}
\usepackage{amssymb}
\usepackage{grffile}
\usepackage[pdftex]{graphicx}
\usepackage{amsmath, amstext, amssymb, amsfonts, amsxtra}
\usepackage{textcomp}
\usepackage{xspace}

\def \ell{{d}}
\def \av#1{{\langle#1\rangle}}

\newcommand{\bra}[1]{\ensuremath{\langle #1 \vert}\xspace}
\newcommand{\ket}[1]{\ensuremath{\vert #1 \rangle}\xspace}
\newcommand{\nv}{\boldsymbol{n}}
\newcommand{\ev}{\boldsymbol{e}}

\newcommand{\bop}{\hat{b}} 
\newcommand{\nop}{\hat{n}} 
\newcommand{\bdop}{\hat{b}^{\dagger}} 
      
\newcommand{\rhom}{\hat{\rho}}

\newcommand*{\bfrac}[2]{\genfrac{}{}{0pt}{1}{#1}{#2}}

\newcommand{\sutd}{Singapore University of Technology and Design, 20 Dover Drive, 138682 Singapore}
\newcommand{\unige}{D\'epartement de Physique Th\'eorique, Universit\'e de Gen\`eve, CH-1211 Gen\`eve, Switzerland.}
\newcommand{\unibonn}{HISKP, Universit\"at Bonn, Nussallee 14-16, D-53115 Bonn, Germany.}
 
\newcommand{\cf}{Coll\`ege de France, 11 place Marcelin Berthelot, 75005 Paris, France.} 
\newcommand{\ep}{Centre de Physique Th\'eorique, Ecole Polytechnique, CNRS, 91128 Palaiseau Cedex, France.}
\newcommand{\dpmc}{DPMC-MaNEP,  Universit\'e de Gen\`eve, CH-1211 Gen\`eve, Switzerland.}

\usepackage{ulem}
\begin{document}

\title{Emergence of glass-like dynamics for dissipative and strongly interacting bosons}                     

\author{Dario Poletti}
\affiliation{\sutd}
\author{Peter Barmettler} 
\affiliation{\unige}
\author{Antoine Georges}
\affiliation{\cf}
\affiliation{\ep}
\affiliation{\dpmc}
\author{Corinna Kollath}
\affiliation{\unige}   
\affiliation{\unibonn}

\begin{abstract}
We study the dynamics of a strongly interacting bosonic quantum gas in an optical lattice potential under the effect of a dissipative environment. We show that the interplay between the dissipative process and the Hamiltonian evolution leads to an unconventional dynamical behavior of local number fluctuations. In particular we show, both analytically and numerically, the emergence of an anomalous diffusive evolution in configuration space at short times and, at long times, an unconventional dynamics dominated by rare events. Such rare events, common in disordered and frustrated systems, are due here to strong interactions. This complex two-stage dynamics reveals information on the level structure of the strongly interacting gas.
\end{abstract}

\pacs{05.70.Ln, 03.75.Kk, 37.10.Jk, 67.85.-d}

\maketitle

Unconventional, non-exponential, relaxation dynamics of a perturbed system towards equilibrium has attracted a lot of interest over decades. Already in 1847, Kohlrausch \cite{Kohlrausch1847} observed a stretched exponential decay in time $t$, i.e.~$e^{-(t/t_0)^{\alpha}}$ with $\alpha\in (0,1)$ and $t_0$ a positive constant, of the discharge of capacitors fabricated from glasses. Since then, such a decay has been observed in many systems such as molecules and polymers \cite{Phillips1996,AngellMartin2000}, spin glasses \cite{ChamberlinOrbach1984,BinderYoung1986}, nano-sized magnetic particles \cite{BedantaKleemann2009}, and certainly amorphous silicon \cite{KakaliosJackson1987, BerthierBiroli2011}.  

A broad variety of theoretical approaches has been developed to explain the mechanism of this unconventional relaxation dynamics \cite{Goldstein1969,BerthierBiroli2011,PalmerAnderson1984,RitortSollich2003}. In many of these approaches, e.g.~the treatment of the Griffiths phase in disordered spin systems \cite{Bray1987}, rare configurations have been identified to play a key role. These configurations have an exponentially small probability to occur, and therefore contribute minimally to the short-time dynamics. However, because their relaxation time scale is very long, these rare configurations can dominate the long time evolution. 
Rare configurations play an important role in the relaxation dynamics of glasses, where they give rise to stretched exponential decays. We will thus refer to this dynamics induced by rare events as `glass-like' in the following. 
 
\begin{figure}[!ht]
 \includegraphics[width=\columnwidth]{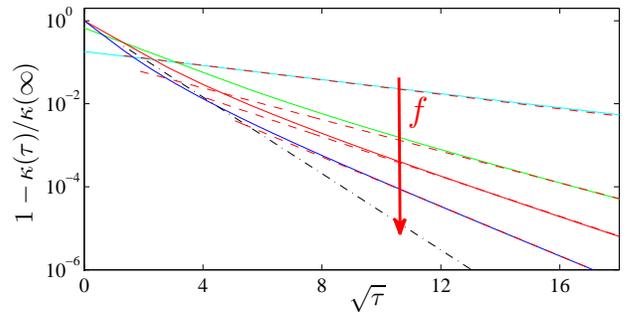}
   \caption{(color online) $1-\kappa(\tau)/\kappa(\infty)$ vs square-root of rescaled time $\tau$ for the interaction over dissipative coupling ratio $U/\hbar\gamma=10$. Numerical results of Eq.~\eqref{eq:evoeq} are shown for different fillings, $f=0.1,0.5,1,3$ in the direction of the arrow, as solid lines and corresponding stretched exponential fits as dashed lines. The analytical result (\ref{eq:compequa}) of the diffusion equation~\eqref{eq:diffequ} is shown as a (black) dot-dashed line.} \label{fig:FigA}      
\end{figure}

In this work, we uncover that also in a quantum many body systems, as the Bose-Hubbard model, the dissipative coupling to a Markovian, i.e.~memory-less, environment can cause glass-like dynamics. We show that the long time behaviour in these systems can be dominated by rare configurations. 
These rare configurations are characterized by a large number of atoms occupying a single lattice site. Increasing the number of atoms on the largely occupied site is associated to a long time scale, since the energetic cost of modifying this kind of configurations is very large. Due to this long time scale, these rare configurations dominate the long time dynamics inducing an unconventional dynamics of stretched exponential form as shown, for the case of local number fluctuations $\kappa=\langle \nop_j^2 \rangle-\langle\nop_j\rangle^2$ (where $\nop_j$ is the number operator of atoms on site $j$) in Fig.~\ref{fig:FigA}. 
Additionally, the glass-like dynamics is preceded by an algebraic relaxation process due to the interplay of many energetically close configurations with low particle fluctuations. 
Therefore, both unconventional dynamics of this open quantum many body system are signatures of the complex structure of its configuration space and energy spectrum that the dissipative term forces to explore. 

The non-exponential decay is in contrast to the typical evolution found for quantum many body systems coupled to a Markovian environment. In these systems, the decay is often dominated by an exponential dynamics, as e.g., the counterintuitive Zeno effect \cite{BrennenWilliams2005,ShchesnovishKonotop2010,BarmettlerKollath2011,SyassenDuerr2008,ZezyulinOtt2012}, or the relaxation to a desirable state driven by an artificially engineered environment \cite{BarreiroBlatt2011,MuellerZoller2011}. Only recently, first signs of an intriguing slowing down of the heating dynamics for interacting bosonic \cite{PichlerZoller2010, PolettiKollath2012, PichlerDaley2012, PichlerZoller2013} and fermionic \cite{BernierKollath2012} gases and an algebraic decay for a specially designed environment which imprints coherence \cite{TomadinZoller2011} have been predicted. However, up to now, the understanding of the large variety of dynamical behaviors that occur in an interacting many body system and of its origin is still a great challenge.

We study the heating of  $N$ ultracold bosonic atoms in an optical lattice of $N_s$ sites with filling $f=N/N_s$ and connectivity (number of nearest neighbors per site) $z$, described by the following master equation \cite{GerbierCastin2010, PichlerZoller2010}:      
\begin{equation}
	\partial_t\hat{\rho}=-\frac{i}{\hbar}[\hat{H},\hat{\rho}]+\mathcal{D}\left(\hat{\rho}\right). \label{eq:master}     
\end{equation}   

The first term describes the unitary evolution of the density matrix $\hat{\rho}$. This evolution is governed, in the single Bloch band limit, by the Bose-Hubbard Hamiltonian $\hat{H}=-J\sum_{\av{j,l}} \bdop_j\bop_l+\frac{U}2 \sum_j \nop_j\left(\nop_j-1\right)$ where $\av{j,l}$ denotes pairs of neighboring sites \cite{JakschZoller1998,BlochZwerger2008}. The operators $\bdop_j$ and $\bop_j$ are bosonic creation and annihilation operators on site $j$ and $\nop_j=\bdop_j\bop_j$ counts the number of atoms. 
The dissipator $\mathcal{D}\left(\hat{\rho}\right)=\gamma \sum_{j}\left(\nop_j\hat{\rho}\nop_j-\frac 1 2 \nop_j^2\hat{\rho}-\frac 1 2 \hat{\rho}\nop_j^2\right)$ models the dissipative coupling to a Markovian environment via the local density with strength $\gamma$. This can be due to a noisy potential both in space and time added to the optical lattice \cite{LyeInguscio2005,LyeInguscio2005b,PichlerDaley2012,PichlerZoller2013}. 
We have restricted the description to the lowest Bloch band of the optical lattice potential. The validity of this approximation is discussed in the conclusions.

\begin{figure}[!ht]
 \includegraphics[width=\columnwidth]{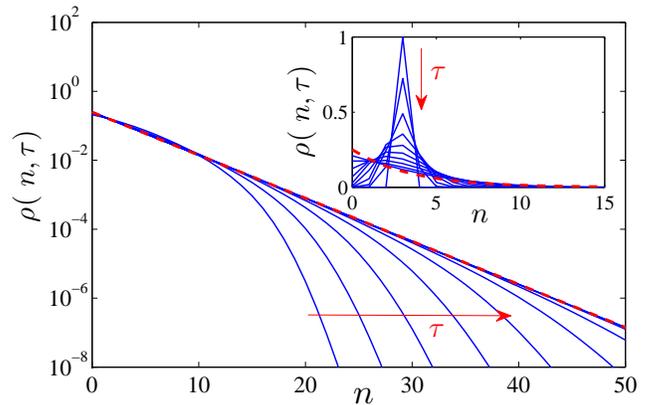}
   \caption{(color online) Numerical evolution of the density matrix elements $\rho(n,\tau)$ (solid lines, Eq.~(2)) in a semi-logarithmic plot, versus $n$ for large rescaled times $\tau$ between 0.1 and 50 (not equidistant) in the direction of the (red) arrow. The inset shows the same evolution at shorter times $\tau$ between 0.0002 and 0.1 (not equidistant) in the direction of the (red) arrow in a linear plot. Parameters: $f=3$, $U/\hbar\gamma=10$. The (red) dashed lines show the (analytical) asymptotic limit. } \label{fig:Fig1}      
\end{figure}

In the following, we study in detail the heating dynamics of a system, initially in its ground state with respect to $\hat H$, under the joint action of dissipation and the Hamiltonian evolution. We concentrate on the strongly interacting regime $U\gg J,\hbar\gamma$. The dissipator causes the off-diagonal elements of the density matrix, in the following always represented in the Fock basis, to decay towards the decoherence free subspace. This consists of all possible diagonal density matrices $\rhom$. 
In the presence of the hopping term, the heating process drives the system to a unique steady state $\rhom(t\!=\!\infty\!)\!=\!\frac {\hat{\mathbb{I}}} M$, the highest entropy state~\cite{PolettiKollath2012}. Here, $M$ is the dimension of the Hilbert space at fixed atom number $N$, and $\hat{\mathbb{I}}$ the identity operator. The approach of this steady state, can be described for $\gamma t \gg 1$, by adiabatically eliminating \cite{BrennenWilliams2005,Garcia-RipollCirac2009} the small off-diagonal elements. A closed set of classical rate equations for the diagonal elements of $\rhom$ is obtained \cite{PolettiKollath2012,PolettiKollath2012b,supp}. The diagonal configurations that are connected are those for which a particle is moved from a site with occupation $m'$ to one of its neighbors with occupation $m$. The process occurs via virtual hopping to and from an off-diagonal element of the density matrix \cite{supp}. To study this dynamics, we use a separable and translationally invariant ansatz $\rhom(t)=\bigotimes_j\left[\sum_{n}\rho(n,t)\ket{n}\bra{n}\right]$ where $j$ runs over all the lattice sites and $n$ over all the possible occupations of each site. The probability distribution $\rho(n,t)$ of the single site occupation evolves as  
\begin{eqnarray}   
  \partial_{\tau} \rho(n,\tau) \!=\!\!\!\!\sum_{m,d=\pm 1} \!\!\!\!\mathcal{T}(n,m,d) \!\!\!&\bigg[&\!\!\!\!\rho(m-d,\tau) \rho(n+d,\tau)  \nonumber \\ 
 &-&   \rho(m,\tau)\rho(n,\tau) \bigg]  \label{eq:evoeq}   
\end{eqnarray}
where $\tau=t/t^*$ with $t^*=\frac{U^2f^2}{2zJ^2\gamma}$ and $\mathcal{T}(m,m',d)=f^2\frac{(m+\delta_{d,1})(m'+\delta_{d,-1})}{(m-m'+d)^2+\left(\hbar\gamma/U\right)^2}$ \cite{supp}. A typical evolution of the occupation number distribution can be acquired by studying Fig.~\ref{fig:Fig1}.
At short times, but still $\gamma t\gtrsim 1$, the very narrow initial distribution around the average filling $f$ broadens almost symmetrically (see inset of Fig.~\ref{fig:Fig1}). After the rapid broadening a new regime with an asymmetric evolution sets in, in which the tail of the distribution slowly converges towards the expected asymptotic distribution $\rho(n,\infty)=\frac 1 f \left(\frac f {1+f}\right)^{n+1}$ \cite{supp}. This means that the probability for states with larger filling is exponentially suppressed, i.e.~these states are rare. Note that $\rho(n,\infty)$ is exactly the single site reduced density matrix of the full asymptotic density matrix $\hat{\rho}(t=\infty)=\frac {\hat{\mathbb{I}}} M $ \cite{Gutzwillernote}.  
To obtain analytical insight into the very different regimes of the evolution, we take the continuum limit of Eq.~(\ref{eq:evoeq}) for large $f$. The continuous on-site occupation number distribution $p(x=n/f,\tau)=f \rho(n,\tau)$ and thus $p((n+1)/f,\tau)=p(x+dx,\tau)=p(x,\tau)+\frac{\partial p}{\partial x}dx$. Hence one derives the non-linear integro-differential equation~\cite{supp} 
\begin{equation} 
\frac{\partial p(x,\tau)}{\partial \tau}=\frac{\partial}{\partial x}\left[D(x,\tau)\frac{\partial p(x,\tau)}{\partial x} - F(x,\tau) p(x,\tau)\right]. \label{eq:diffequ}           
\end{equation} 
Here 
\begin{equation} 
D=\int_0^{\infty} \frac{xy \;\; p(y,\tau) }{(x-y)^2+\varepsilon^2}  dy ,\;\;\; 
F=\int_0^{\infty} \frac{xy \;\; \partial_y p(y,\tau)}{(x-y)^2+\varepsilon^2}  dy \label{eq:DF}
\end{equation} 
and $\varepsilon=\hbar\gamma/fU$. The peculiar form of $D(x,\tau)$ and $F(x,\tau)$ stems from the configuration dependent rates and triggers a wide range of rich phenomena. Note that the structure of (\ref{eq:DF}) ensures that both the total probability ($\int_0^{\infty}p(x,\tau)dx=1$) and the average population ($\int_0^{\infty}x\;p(x,\tau)dx=1$) are conserved quantities \cite{supp}. Further, it can be checked that the asymptotic solution of (\ref{eq:diffequ}) is $p(x,\infty)= e^{-x}$, which is the continuum limit of the steady state in the large-$f$ limit \cite{supp}.  
The continuum description is justified for $f$ large and finite $\varepsilon$, assuming that $p(x,\tau)$ varies smoothly enough on scales of the order of $1/f$. In the present case, the strongest variations of distributions are due to the initial state, especially for a low filling $f$. After this initial stage $p(x,\tau)$ smoothens out rather quickly and the continuum description is highly accurate over a wide time range. 

In the following we solve analytically the diffusion equation (\ref{eq:diffequ}) in the short time and in the long time limits focusing on the evolution of the particle distribution and the local density fluctuations $\kappa$.

{\it Short time relaxation:}
Within the diffusion equation (\ref{eq:diffequ}), initially the distribution $p$ is strongly peaked and symmetric around the value $x=1$. For such a distribution the force is negligible compared to the diffusion function. The diffusion equation at $x\approx 1$ can be approximated by $\partial_{\tau}p(x,\tau)= \partial_x \left[\frac{1}{(x-1)^2+\varepsilon^2} \partial_x p(x,\tau) \right]$.  

This leads to a dynamics which, in the analytically solvable limit $\varepsilon\!\rightarrow\! 0$, is given by an anomalous diffusion of the form $p(x,\tau)= \frac{1}{4\;\Gamma(5/4)\;\tau^{1/4}}  e^{-\frac{(x-1)^4}{16\tau}}$. $\Gamma(s)$ is the gamma function \cite{AbramowitzStegun1972}. 
\begin{figure}[!ht]
 \includegraphics[width=\columnwidth]{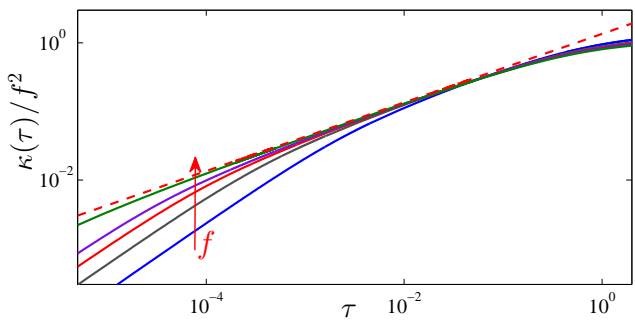}
   \caption{(color online) Local particle fluctuations $\kappa/f^2$ versus rescaled time $\tau$ for $U/\hbar\gamma=10$. Numerical results (solid lines) for various fillings, $f=3,5,7,9,20$ in the direction of the arrow, are obtained solving Eq.~\eqref{eq:evoeq}. The approximate analytical solution $\kappa/f^2=\frac{\Gamma(3/4)}{\Gamma(5/4)}\sqrt{\tau}$ of the diffusion equation is represented by the dashed line.} \label{fig:compressibility}      
\end{figure}
Using this analytical solution for the particle distribution, the local number fluctuations ${\kappa(\tau)}/{f^2}=\int_0^{\infty} (x^2-1) p(x,\tau)dx$ exhibit a power-law relaxation with $\kappa/f^2=\frac{\Gamma(3/4)}{\Gamma(5/4)}\sqrt{\tau}$. This analytical result is in excellent agreement with the numerical results shown in Fig.~\ref{fig:compressibility} obtained by solving Eq.~(\ref{eq:evoeq}). Deviations are found at small fillings and short times, where the approximation $\varepsilon\to 0$ is not justified. However, for larger fillings, for example $f=20$, the time-regime in which the power-law decay appears is already large. 

Physically, this very rapid initial broadening of the particle distribution translates into the fast creation of small particle fluctuations around the average value caused by the heating. These fluctuations arise via virtual excitations of low energetic cost of order O(U) which thus can be reached rapidly. This behavior is similar to the dynamics observed in a double well potential \cite{PolettiKollath2012}.

{\it Long time relaxation:}
The obtained short time solution breaks down as the distribution approaches the reflective boundary at $x=0$ \cite{Boundary}; the distribution is no longer symmetric around $x=1$, and the combined action of the force term with the diffusion drives the system towards its large time asymptotics $p(x,\infty)=e^{-x}$. Physically, the exponential suppression of large values of $x$ corresponds to the rareness of the states with a high number of particles accumulated on a single site. Therefore, naively one expects that their effect is overwhelmed by the much more numerous states at low filling. However, the rare states are associated with a decaying small diffusion function and force given by $D(x,\tau)\approx -F(x,\tau) \approx \frac 1 x $ leading to the slow occupation of the states with large $x$. Due to these large time scales, these rare states are found to dominate the long time dynamics despite their exponentially suppressed probability to occur. The underlying quantum mechanical process behind this slow diffusion is the large energy cost of the virtual states via which the diffusion processes at high $x$ take place.    

The form of the forcing term $F$ in the large $x$ limit brings connections to other intriguing physical problems. One example is the emergence of non-ergodic and super-aging behavior for diffusion in a logarithmic potential, but with a constant diffusion function $D$ \cite{KesslerBarkai2010}. This can be realized in another category of experiments with dissipative cold atoms in optical lattices \cite{CastinCohenTannoudji1991,DouglasRenzoni2006}. 

\begin{figure}[!ht]
 \includegraphics[width=\columnwidth]{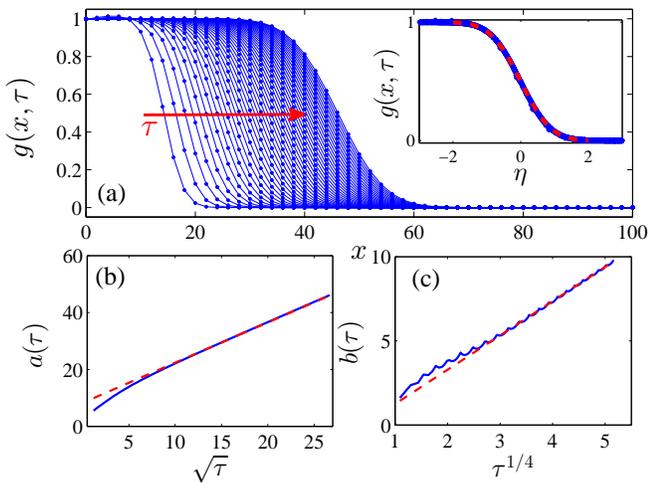}
   \caption{(color online) (a) The blue-dotted line shows the evolution of $g(x,\tau)$ versus $x$ 
   for large rescaled times $\tau\in [28,710]$ in the direction of the red arrow. Inset: Plot of $g(x,\tau)$ versus $\eta=(x-a(\tau))/b(\tau)$ (blue line) and fit with an error function (red dashed line). (b) $a(\tau)$ is plotted versus $\sqrt{\tau}$ (blue line) and compared to a linear fit (red dashed line). 
   (c) $b(\tau)$ is plotted versus $\tau^{1/4}$ (blue line) and compared to a linear fit (red dashed line). 
   Parameters: $f=0.5$, $U/\hbar\gamma=15$.} \label{fig:Fig2}      
\end{figure}

Approaching the asymptotic solution (see Fig.~\ref{fig:Fig1}), we use the convenient ansatz $p(x,\tau)=p(x,\infty)g(x,\tau)$. 
The evolution of the function $g$ is shown in Fig.~\ref{fig:Fig2} and suggests a scaling form $g(x,\tau)=g(\eta)$ with $\eta=\!\frac{x-a(\tau)}{b(\tau)}$. Here $a$ and $b$ are some functions of $\tau$ to be determined. 
Adopting this ansatz leads to \cite{supp}  
\begin{equation}
 p(x,\tau)= \frac{p(x,\infty)} 2 \; \left\{1-{\rm erf}\left[ \frac{\sqrt{3}}{2}\; \frac{x-\sqrt{2\tau}}{(2\tau)^{1/4}} \right] \right\}  \label{eq:erf}   
\end{equation}
where ${\rm erf}$ is the error function \cite{AbramowitzStegun1972}. 

Figs.~\ref{fig:Fig2}(b-c) show that, at long enough times, the numerical data and the proposed analytical $\tau$-dependence $a(\tau)\propto \sqrt{\tau}$ and $b(\tau)\propto \tau^{1/4}$ match accurately. Note that the numerical results still show deviations from the exact analytical prefactors. We verified that these deviations become smaller with increasing filling. Thus we conclude that the analytical form is applicable in the large $f$ limit. For large enough times, the obtained solution gives a very good approximation in the entire range of $x$. The reason for this is the fast initial relaxation at low values of $x$. Thereafter, only small relative changes occur at small $x$. These changes in the probability distribution are mainly connected to the variations at large $x$ via particle number conservation. 
Thus, we can use the obtained solution to calculate the local particle fluctuations $\kappa$. The corresponding integral can be solved analytically giving   
\begin{equation}
 \kappa_{\infty}-\kappa(\tau)\propto h(\tau) \;e^{-\frac 3 2 \sqrt{\frac{\tau}{2}}} \label{eq:compequa}
\end{equation}
where $h$ depends algebraically on $\tau$. We thus have shown analytically the emergence of the stretched exponential behavior. This finding, as depicted in Fig.~\ref{fig:FigA}, compares well to the numerical solution of Eq.~(\ref{eq:evoeq}). However, the time at which the stretched exponential occurs, and the detailed decay, depend on the filling. In particular, the stretched exponential occurs later for larger fillings. 

To summarize, we have uncovered, in the Bose-Hubbard model coupled to a dissipative environment, two unconventional relaxation regimes: at short times, and large enough fillings, a power-law regime, while at large times, and any filling, a stretched exponential regime. This last regime is dominated by rare events which correspond to the occupation of a single site with a large number of atoms. The rare states in the tail of the distribution function $e^{-x}$ are occupied with decreasing time scales $\propto 1/x$. These ingredients alone allow to estimate the main time dependence of the fluctuations employing a simple saddle point argument. Since $\kappa\approx \int_0^{\infty}x^2 e^{-x}e^{-\frac{At}{x}}$, the saddle point integration $\left( \frac{d}{dx}\left( x+\frac{At}{x} \right)|_{x_0}=0\right)$ leads to $ \kappa\approx e^{-\sqrt{At}} $ recovering the stretched exponential. Indeed, very slow transition rates, due to the high energetic cost of the processes connecting these rare configurations, dominate the long time dynamics. This emergent glass-like dynamics is thus a signature of the complex level structure of the Bose-Hubbard Hamiltonian. 
Dissipation, forcing the system to explore its whole configuration space, including rare and energetically unfavorable configurations, manifests the complex energy levels structure of the system. 
In future works, we plan to investigate existence of stronger connections beyond the dominating rare events to glassy physics, e.g.~the emergence of dynamical heterogeneity, aging phenomena \cite{BerthierBiroli2011,NussinovBalatsky2012} or the physics of kinetic constraints model \cite{RitortSollich2003,OlmosGarragan2012}. 
   
Experimental observation of these relaxation regimes is within reach. We discuss in the following possible realizations for the stretched exponential regime which is experimentally more demanding to study. 
We consider a gas of $^{87}$Rb atoms (mass $m$) confined to an optical lattice potential with wavelength $\lambda = 1064nm$. For a lattice depth of $V=9\; E_R$ (where $E_R=h^2/2m\lambda^2$ ), the ratio of the interaction over tunneling is $U/J\approx 9.2$ with $J/\hbar\approx 367\;s^{-1}$. 
The realization of the dissipator in Eq.(1) could be achieved by a noisy optical potential pattern, e.g.~due to an additional, time dependent, speckle beam \cite{LyeInguscio2005} or an incommensurate superlattice \cite{LyeInguscio2005b} with a randomly changing phase or amplitude. 
Since the strength of $\gamma$ in these setups can be tuned by the intensity of the light fields, this allows to have $\hbar\gamma/J\approx 1$ or more. 
To observe the stretched exponential regime, a low filling, for example $n=0.5$ would be advantageous (see Fig.\ref{fig:FigA}). For this filling and $\hbar\gamma/J\approx 1$, the experimental time-scales needed to identify the stretched exponential regime are of the order of $t>0.3$s ($\sqrt{\tau}>8$ in Fig.~\ref{fig:FigA}) and lattice occupations up to $4$ atoms per site will be occupied with a non-negligible probability. This required time-scale is small compared to the time-scale of the three-body losses (for occupation of $n=4$, the three-body loss scale is approx.~$1.6$s \cite{CampbellKetterle2006}) and to the time-scale of secondary collisions with the background gas \cite{CampbellKetterle2006}. Thus these processes can be neglected.  
Further, transitions to higher Bloch bands can be due to interaction of the highest occupied sites or by the dissipation. Both can be neglected for experimentally relevant time-scales as i) the next Bloch band is approximately at an energy $\Delta E\approx \sqrt{4V E_R}$ which is $\approx 3.7$ times the interaction energy $\frac{U}{2}n(n-1)$ for a large filling as $n=4$ and ii) transitions due to dissipation can be controlled via engineering the noise spectrum and tuning the Lamb-Dicke parameter \cite{TransitionsNote,PichlerDaley2012, PichlerZoller2013}. For example the frequency of the noise pattern could be cut below the frequency corresponding to transitions to higher bands.

We are grateful to J.S.~Bernier, G.~Biroli, R. Bouffanais, J.-P. Eckmann, J.B. Gong, P.~H\"anggi, H.~Ott, T.~Prosen, and P. Wittwer for fruitful discussions. 
We acknowledge ANR (FAMOUS), SNSF (Division II, MaNEP), SUTD start-up grant (SRG-EPD-2012-045) and the DARPA-OLE program for financial support.    

{\it Note added.} During the referee process we became aware of \cite{CaiBarthel}, which shows non-exponential relaxation dynamics.

\newpage

\appendix*
\section{Supplementary material}

\paragraph{Adiabatic elimination -- } In order to derive the effective equations for diagonal elements of the density matrix (Eq. (2) of the main text), we represent $\hat{\rho}$ in the Fock basis as $\hat{\rho}=\sum_{\nv,\nv'}\rho^{\nv}_{\nv'}\ket{\nv}\bra{\nv'}$. The $\nv$ and $\nv'$ are vectors that describe the distribution of the atoms on the lattice sites, i.e. $\nv=(..,n_l,..)$ where $n_l$ is the number of atoms on site $l$. This representation is advantageous, since the decoherence free subspace of the dissipator corresponds to the density matrices with only nonvanishing diagonal entries. This decoherence free subspace dominates the long-time dynamics. Quantum fluctuations around the decoherence free subspace can be included perturbatively in the hopping term using the adiabatic elimination method \cite{Garcia-RipollCirac2009,PolettiKollath2012,PolettiKollath2012b}. This leads, to second order in the kinetic term of the Hamiltonian, to a virtual tunneling process of a particle to the neighbouring site, then it dissipates (at a rate $\gamma$) or dephases (due to the interaction $U$), and eventually tunnels another time. 

For example, for the case of the tunneling of an atom from site $l'$ with $n_{l'}$ atoms to site $l$ with $n_l$ atoms, we can compute the time dependence of the off-diagonal terms by integrating 
\begin{align}   
  i\hbar\partial_t \rho^{\nv+\ev^1_{l,l'}}_{\nv}(t) &= \left[ -i\hbar\gamma + (n_l-n_{l'}+1)U  \right] \rho^{\nv+\ev^1_{l,l'}}_{\nv}(t) \nonumber  \\    
      &-  J\sqrt{(n_l+1)n_{l'}} \left[ \rho^{\nv}_{\nv}(t)  - \rho^{\nv+\ev^1_{l,l'}}_{\nv+\ev^1_{l,l'}}(t) \right],  \tag{1S}  
\end{align} 
which is derived from Eq.~(1) of the main paper. Here we have used the vector $\ev^d_{l,l'}$ which is $0$ everywhere except at the neighboring positions $l$ and $l'$ where it takes respectively the values $d$ and $-d$. The vector $\ev^d_{l,l'}$ connects two configurations which are related just by the tunneling of a single atom, the only relevant dynamical process that can change the distribution of the atoms in the lattice.
The integration of Eq. (1S) is well approximated by [2,3]     
\begin{align}   
 \rho^{\nv+\ev^1_{l,l'}}_{\nv}(t) &=- \frac{J\sqrt{(n_l+1)n_{l'}}\left[  \rho^{\nv}_{\nv}(t)  - \rho^{\nv+\ev^1_{l,l'}}_{\nv+\ev^1_{l,l'}}(t)   \right]}{\left[(n_l-n_{l'}+1)U+i\hbar\gamma\right]},  \tag{2S} \label{eq:off-diago}            
\end{align}
since $\hbar\gamma$ or $U$ are much larger than $J$. The off-diagonal elements can thus be described only by the diagonal ones.

When the number of atoms at site $l$ changes by $d$ and those at the neighboring site $l'$ by $-d$, the evolution of the diagonal terms of the density matrix, belonging to the dissipation-free space, is derived using Eq.~(\ref{eq:off-diago}) and Eq.~(1). This results in the following set of equations:  
%
\begin{equation}    
 \partial_{\tau}\,\rho^{\nv}_{\nv}(\tau)\! =\! -\!\!\!\sum_{\bfrac{\av{l,l'}}{d=\pm 1}}\!\!\frac 1 z \mathcal{T}\!\left(n_{l},n_{l'},d\right) \left( \rho^{\nv}_{\nv}(\tau) -\rho^{\nv+\ev^d_{l,l'}}_{\nv+\ev^d_{l,l'}}(\tau) \right), \tag{3S}
\label{eq:evodiag}       
\end{equation} 
where $\mathcal{T}(m,m',d)=f^2\frac{(m+\delta_{d,1})(m'+\delta_{d,-1})}{(m-m'+d)^2+\left(\hbar\gamma/U\right)^2}$, $f$ is the filling, $z$ is the number of nearest neighbors, $\tau=t/t^*$ with $t^*=\frac{U^2f^2}{2zJ^2\gamma}$. 
In the thermodynamic limit $N\rightarrow\infty$, keeping the filling $f=N/N_s$ constant, 
one can separate the diagonal density matrix
$\rhom(\tau)=\bigotimes_j\left[\sum_{n}\rho_j(n,\tau)\ket{n}\bra{n}\right]$. 
Using translation invariance, i.e. $\rho(n,\tau)=\rho_j(n,\tau)$ for all $j$, one
obtains a non-linear equation for the reduced single site
density matrix which is Eq.(2) of the main text.\\ 

\paragraph{Steady state --} In Ref. \cite{PolettiKollath2012}, the steady state density matrix $\rhom(\tau=\infty)=\sum_{\nv}\frac 1 M |\nv\rangle\langle\nv| $ had been identified. In order to obtain the one-site reduced density matrix elements $\rho_l(n_l,\infty)=\sum_{n_j,j\neq l}\rho_{\nv}^{\nv}(\infty)$, the number of possible configurations on a lattice needs to be taken into account. On a generic lattice, the number of configurations in which $N$ atoms can be distributed between $N_s$ sites is given by the binomial coefficient $M=\left(\bfrac{N_s+N-1}{N}\right)$. $M$ corresponds to the dimension of the Hilbert space. Thus, the one-site reduced density matrix elements are   
\begin{equation}
 \rho(n,\infty) = \frac 1 M\left(\bfrac{N_s+N-n-2}{N-n}\right) \tag{4S}  \,.  
\end{equation}
where, given translation invariance, we have simplified the notation using $\rho(n,\infty)\equiv\rho_l(n_l,\infty)$. 
In the limit $N_s\rightarrow\infty$ and $N/N_s=f$ with $f$ (the filling of the lattice) kept constant, using Stirling's approximation $n!\approx \sqrt{2\pi n}\left(\frac n e\right)^n$, leads to   
\begin{equation}
\rho(n,\infty) \approx \frac 1 f \left(\frac{f}{1+f}\right)^{n+1}\!\!\!\!. \tag{5S} \label{eq:steadys}    
\end{equation}
This result is independent of the dimensionality of the system. Introducing a rescaled parametrization for the occupation, $x=n/f$, we obtain, in the limit of large fillings and using $\lim_{f\rightarrow\infty}\left(1+\frac 1 f\right)^{fx}=e^x$, that  
\begin{equation}
p(x=n/f,\tau=\infty)=\lim_{f\rightarrow\infty}f\rho(n,\infty)=e^{-x}. \tag{6S} \label{eq:steady2s}    
\end{equation}  
This shows that the probability for configurations with a large occupation on a single site are exponentially suppressed.\\ 

\paragraph{Derivation of the evolution equation in the continuum limit --} From Eq.~(2) of the main paper, using $p(x=n/f,\tau)=f \rho(n,\tau)$ and $p((n+1)/f,\tau)=p(x+dx,\tau)$, where $1/f=dx$, it is possible to derive the evolution equation in the continuum limit. Taking $\mathcal{T}_c(x,y)=\frac{xy}{(x-y)^2+\epsilon^2}$ we can rewrite Eq.(2) as    
%
%
\begin{eqnarray}   
  \partial_{\tau}&&  \!\!\!\!\!p(x,\tau)=\!\!\!\!\sum_{\nu=\pm 1}\!\!f^2\!\!\int\!\! dy \mathcal{T}_c(x+\delta_{\nu,1}/f,y+\delta_{\nu,-1}/f) \nonumber   \\ 
&&\!\!\!\times\bigg[p(y-\nu dy,\tau) p(x+\nu dx,\tau) -  p(y,\tau)p(x,\tau) \bigg]  \nonumber  \label{eq:evoeqsupp}   
\end{eqnarray}
which expanded to second order gives        
\begin{align}
\partial_{\tau} p(x,\tau)&=\int_0^{\infty}  \bigg\{ \left(\frac{\partial \mathcal{T}_c(x,y)}{\partial y}-\frac{\partial\mathcal{T}_c(x,y)}{\partial x}\right) \times \nonumber \\ 
&  \left[ \frac{\partial p(y,\tau)}{\partial y} p(x,\tau) - \frac{\partial p(x,\tau)}{\partial x} p(y,\tau)  \right] \nonumber \\ 
& +\left[ \frac{\partial^2p(y)}{\partial y^2}p(x)+ \frac{\partial^2p(x)}{\partial x^2}p(y) - 2\frac{\partial p(y)}{\partial y}\frac{\partial p(x)}{\partial x}  \right] \nonumber \\  
& \times \mathcal{T}_c(x,y) \bigg\} dy   \nonumber          
\end{align}
This can be rewritten in a more compact and elegant form 
\begin{eqnarray}
\partial_{\tau}p(x,\tau)=\!\!\int_0^{\infty}\!\!\!\!\!\! \left(\partial_y\!-\!\partial_x\right)\left[\mathcal{T}_c(x,y) \left(\partial_y\!-\!\partial_x\right)p(x,\tau)p(y,\tau)\right] dy   \nonumber          
\end{eqnarray}
and after some manipulation takes the form of the diffusion equation (3).\\

\paragraph{Properties of the evolution equation in the continuum limit -- } From the definition of $p(x,\tau)$ it follows that the probability to find a certain configuration is normalized to unity, i.e.~$||p||:=\int_0^{\infty}p(x,\tau)dx=1$ and it is constant in time. Further, the average rescaled filling $\langle x\rangle=\int_0^{\infty}x \;p(x,\tau)dx=1$ is also constant in time, which corresponds to the conservation of particles. The conservation of both can be shown explicitely: 
\begin{align}
\frac{d ||p||}{d\tau}&=\int_0^{\infty}\frac{\partial p}{\partial \tau} dx \tag{7S} \\    
&=\left[D(x,\tau)\frac{\partial p(x,\tau)}{\partial x} - F(x,\tau) p(x,\tau)\right]_0^{\infty}=0 \nonumber
\end{align} 
and 
\begin{align}
&\frac{d \langle x \rangle}{d\tau}=\int_0^{\infty}x\frac{\partial p}{\partial \tau} dx  \tag{8S} \\    
&=\int_0^{\infty}x \left\{\frac{\partial}{\partial x}\left[D(x,\tau)\frac{\partial p(x,\tau)}{\partial x} - F(x,\tau) p(x,\tau)\right]\right\}   dx  \nonumber \\    
&=x\left[D(x,\tau)\frac{\partial p(x,\tau)}{\partial x} - F(x,\tau) p(x,\tau)\right]_0^{\infty} \nonumber \\ 
&-\!\!\int_0^{\infty}\!\! \left[D(x,\tau)\frac{\partial p(x,\tau)}{\partial x} - F(x,\tau) p(x,\tau)\right]dx \nonumber \\
&=-\!\!\int\!\!\!\!\int_0^{\infty}\!\! \left\{\frac{xy\left[p(y,\tau)\partial_x p(x,\tau) - \partial_y p(y,\tau) p(x,\tau) \right]}{(x-y)^2+\epsilon^2}\right\}dxdy \nonumber \\ 
& =0 \nonumber   
\end{align} 
where we have used the definitions of $D(x,\tau)$ and $F(x,\tau)$ given in the main paper (Eq.~(4)).\\ 

\paragraph{ Solution of the continuum equation for large times --} By using the ansatz $p(x,\tau)=p(x,\infty)g(x,\tau)$ and the scaling $g(x,\tau)=g\left(\frac{x-a(\tau)}{b(\tau)}\right)$, Eq.~(3) reduces to  
\begin{equation}
\left[b\left(\dot{a}a-1\right)+\eta\left( ab\dot{b}+b^2\dot{a} \right) \right]\frac {d g} {d \eta} + \frac {d^2 g} {d \eta^2} = 0 \tag{9S} \label{eq:diffequg}
\end{equation}
where the dot stands for the derivative with respect to $\tau$ and $\eta=\frac{x-a(\tau)}{b(\tau)}$. Since $g$ is only a function of $\eta$ then 
%
%
\begin{align} 
&b(\dot{a}a-1)=\alpha \tag{10S} \\ 
& ab\dot{b}+b^2\dot{a} =\beta \tag{11S} 
\end{align}
where $\alpha$ and $\beta$ are two constants. The only physically relevant solution is such that $a(\tau)=\sqrt{2\tau}$ and $b=\sqrt{\frac 2 3 \beta\sqrt{2\tau}}$. 
Using this insight, Eq.~(\ref{eq:diffequg}) becomes 
\begin{equation}
\beta\eta\frac {d g} {d \eta} + \frac {d^2 g} {d \eta^2} = 0 \tag{12S} \label{eq:diffequg2}
\end{equation}
which has the solution 
\begin{equation}
g(\eta) = c_1 + c_2 \;{\rm erf}\!\left(\sqrt{\frac {\beta} 2} \eta\right) \tag{13S} \label{eq:gerf}
\end{equation}
Given the boundary conditions, $g(0)=1$ and $g(\infty)=0$, we obtain Eq.~(5).
%



\begin{thebibliography}{10} 

\bibitem{Kohlrausch1847}
R. Kohlrausch, Ann. Phys. (Leipzig) {\bf 12},  393  (1847).

\bibitem{Phillips1996}
J.~C. Phillips, Reports on Progress in Physics {\bf 59},  1133  (1996).

\bibitem{AngellMartin2000}
C.~A. Angell {\it et~al.}, Journal of Applied Physics {\bf 88},  3113  (2000).

\bibitem{ChamberlinOrbach1984}
R.V. Chamberlin, G. Mozurkewich, and R. Orbach, Phys.~Rev.~Lett. {\bf 52},  867
  (1984).

\bibitem{BinderYoung1986}
K. Binder and A.~P. Young, Rev. Mod. Phys. {\bf 58},  801  (1986).

\bibitem{BedantaKleemann2009}
S. Bedanta and W. Kleemann, Journal of Physics D: Applied Physics {\bf 42},
  013001  (2009).

\bibitem{KakaliosJackson1987}
J. Kakalios, R.A. Street, and W.B. Jackson, Phys.~Rev.~Lett. {\bf 59},  1037
  (1987).

\bibitem{BerthierBiroli2011}
L. Berthier and G. Biroli, Review of Modern Physics {\bf 83},  587  (2011).

\bibitem{Goldstein1969}
M. Goldstein, J.~Chem.~Phys. {\bf 51},  3728  (1969).

\bibitem{PalmerAnderson1984}
R.G. Palmer, D.L. Stein, E. Abrahams, and P.W. Anderson, Physical Review Letters {\bf
  53},  958  (1984).

\bibitem{RitortSollich2003}
F. Ritort and P. Sollich, Advances in Physics {\bf 52},  219  (2003).

\bibitem{Bray1987}
A.J. Bray, Phys.~Rev.~Lett. {\bf 59},  586  (1987). 

\bibitem{BrennenWilliams2005} 
G.K. Brennen, G. Pupillo, A.M. Rey, C.W. Clark, C.J. Williams, Journal of Physics B {\bf 38}, 1687 (2005)

\bibitem{ShchesnovishKonotop2010}
V.~S. Shchesnovich and V.~V. Konotop, Phys. Rev. A {\bf 81},  053611  (2010).

\bibitem{BarmettlerKollath2011}
P. Barmettler and C. Kollath, Phys. Rev. A {\bf 84},  041606  (2011).

\bibitem{SyassenDuerr2008}
N. Syassen {\it et~al.}, Science {\bf 320},  1329  (2008).

\bibitem{ZezyulinOtt2012}
D.~A. Zezyulin, V.V. Konotop, G. Barontini, and H. Ott, Phys.~Rev.~Lett. {\bf
  109},  020405  (2012).

\bibitem{BarreiroBlatt2011}
J.~T. Barreiro {\it et~al.}, Nature {\bf 470},  486  (2011).

\bibitem{MuellerZoller2011}
M. Mueller {\it et~al.}, New Journal of Physics {\bf 13},  085007  (2011).

\bibitem{PichlerZoller2010}
H. Pichler, A.~J. Daley, and P. Zoller, Phys. Rev. A {\bf 82},  063605  (2010).

\bibitem{PolettiKollath2012}
D. Poletti, J.-S. Bernier, A. Georges, and C. Kollath, Physical Review Letters
  {\bf 109},  045302  (2012).

\bibitem{PichlerDaley2012} 
H. Pichler, J. Schachenmayer, J. Simon, P. Zoller, A.J. Daley, Phys. Rev. A {\bf 86}, 051605(R) (2012)

\bibitem{PichlerZoller2013} 
H. Pichler, J. Schachenmayer, A.J. Daley, P. Zoller, Phys. Rev. A {\bf 87}, 033606 (2013)

\bibitem{BernierKollath2012}
J.-S. Bernier, P. Barmettler, D. Poletti, and C. Kollath, Phys. Rev. A {\bf 87}, 063608 (2013) 

\bibitem{TomadinZoller2011}
A. Tomadin, S. Diehl, and P. Zoller, Phys. Rev. A {\bf 83},  013611  (2011).

\bibitem{GerbierCastin2010}
F. Gerbier and Y. Castin, Phys.~ Rev.~ A {\bf 82},  013615  (2010).

\bibitem{JakschZoller1998}
D. Jaksch, C. Bruder, J.I. Cirac, C.W. Gardiner,P. Zoller, Phys.~ Rev.~ Lett. {\bf 81},  3108  (1998).

\bibitem{BlochZwerger2008}
I. Bloch, J. Dalibard, and W. Zwerger, Rev.~Mod.~Phys. {\bf 80},  885  (2008).

\bibitem{LyeInguscio2005} 
J.E. Lye, L. Fallani, M. Modugno, D.S. Wiersma, C. Fort, M. Inguscio, Phys. Rev. Lett. {\bf 95}, 070401 (2005)

\bibitem{LyeInguscio2005b} 
E. Lye, L. Fallani, C. Fort, V. Guarrera, M. Modugno, D. S. Wiersma, M. Inguscio, Phys. Rev. A {\bf 75}, 061603(R) (2007)      

\bibitem{Garcia-RipollCirac2009}
J.~J. Garc{\'i}a-Ripoll {\it et~al.}, New Journal of Physics {\bf 11},  013053
  (2009).

\bibitem{PolettiKollath2012b}
D. Poletti, J.-S. Bernier, A. Georges, and C. Kollath, in {\it Proceedings of the Annual International Conference on Optoelectronics, Photonics and Applied Physics}, (GSTF, Singapore, 2013).

\bibitem{supp} For more detailed information refer to the supplementary material. 

\bibitem{Gutzwillernote}
This stands in contrast to steady states which one finds using a Gutzwiller-like approximation from the start \cite{PichlerZoller2010}, since there the suppression of the single particle correlations ``freezes'' the system in many distinct states. 

\bibitem{AbramowitzStegun1972}
M. Abramowitz and I.~A. Stegun, {\it Handbook of mathematical functions with
  Formulas, Graphs, and Mathematical Tables} (New York: Dover Publications,
  USA, 1960).

\bibitem{Boundary} 
At $x=0$ the reflective boundary condition is $\int_0^{\infty}\!\! \textrm{d}y \; \frac{y}{y^2+\varepsilon^2} \left[ \partial_y p(y,\tau)p(0,\tau) -p(y,\tau)\partial_xp(x,\tau)|_{x=0}\right]\!=\!0.$ 

\bibitem{KesslerBarkai2010}
D.A. Kessler and E. Barkai, Phys.~Rev.~Lett. {\bf 105},  120602  (2010).

\bibitem{CastinCohenTannoudji1991}
Y. Castin, J. Dalibard, and C. Cohen-Tannoudji, {\it Light Induced Kinetic
  Effects on Atoms, Ions and Molecules} (L. Moi et al. ETS Editrice, Pisa,
  1991).

\bibitem{DouglasRenzoni2006}
P. Douglas, S. Bergamini, and F. Renzoni, Phys.~Rev.~Lett. {\bf 96},  110601 (2006).

\bibitem{NussinovBalatsky2012} 
Z. Nussinov, P. Johnson, M.J. Graf and A.V. Balatsky, Phys. Rev. B {\bf 87}, 184202(2013). 

\bibitem{OlmosGarragan2012}
B. Olmos, I. Lesanovsky, and J.P. Garrahan, Phys.~Rev.~Lett. {\bf 109},  020403 (2012). 



\bibitem{CampbellKetterle2006} 
G.K. Campbell, J. Mun, M. Boyd, P. Medley, A.E. Leanhardt, L. Marcassa, D.E. Pritchard, W. Ketterle, Science {\bf 313} , 649-652 (2006). 

\bibitem{TransitionsNote} 
Transitions to higher Bloch bands would occur with a rate $\xi=\eta^4 S_2$ \cite{PichlerDaley2012, PichlerZoller2013} where $\eta=\left(\frac{E_R}{4V}\right)^{1/4}$ is the Lamb-Dicke parameter and $S_2$ is the noise spectrum at for the transition between the lowest and the second excited Bloch band. Both parameters can be tuned such that $\xi$ is, at most, of the same order of three body losses.  

\bibitem{CaiBarthel} Z. Cai and T. Barthel, Phys. Rev. Lett. {\bf 111}, 150403 (2013).


\end{thebibliography}
\end{document}